\documentclass[superscriptaddress,amssymb,amsmath,twocolumn]{revtex4}

\usepackage{amsmath}
\usepackage{amssymb}
\usepackage{mathrsfs}
\usepackage{graphicx}
\usepackage{multirow}
\usepackage{array}

\begin{document}


\title{Flow-induced pruning of branched systems\\
and brittle reconfiguration}

\author{Diego~Lopez}
\email{lopez@ladhyx.polytechnique.fr}
\affiliation{Department of Mechanics, LadHyX  Ecole Polytechnique-CNRS, 91128 Palaiseau, France}
\author{S\'ebastien~Michelin}
\email{sebastien.michelin@ladhyx.polytechnique.fr}
\affiliation{Department of Mechanics, LadHyX  Ecole Polytechnique-CNRS, 91128 Palaiseau, France}
\author{Emmanuel~de~Langre}
\email{delangre@ladhyx.polytechnique.fr}
\affiliation{Department of Mechanics, LadHyX  Ecole Polytechnique-CNRS, 91128 Palaiseau, France}
\date{\today}

\begin{abstract}

Whereas most plants are flexible structures that undergo large deformations under flow, another process can occur when the plant is broken by heavy fluid-loading. We investigate here the mechanism of such possible breakage, focusing on the flow-induced pruning that can be observed in plants or aquatic vegetation when parts of the structure break under flow. By computation on an actual tree geometry, a 20-yr-old walnut tree (\textit{Juglans Regia L.}) and comparison with simple models, we analyze the influence of geometrical and physical parameters on the occurrence of branch breakage and on the successive breaking events occurring in a tree-like structure when the flow velocity is increased. We show that both the branching pattern and the slenderness exponent, defining the branch taper, play a major role in the breakage scenario. We identify a criterion for branch breakage to occur before breakage of the trunk. In that case, we show that the successive breakage of peripheral branches allows the plant to sustain higher flow forces. This mechanism is therefore similar to elastic reconfiguration, and can be seen as a second strategy to overcome critical events, possibly a widespread solution in plants and benthic organisms.

\end{abstract}


\maketitle
\section{Introduction}
\label{sec:Introduction}

Most living systems are surrounded by a fluid, be it air or water. When this fluid flows, it generates mechanical forces, that may have major consequences on growth as well as on reproduction or survival \cite{moulia_2006,koehl_2008,delangre_2008}. Typical cases are trees subjected to wind or corals subjected to water currents. In terms of flow-induced deformations, two typical behaviors can be pointed out. In the most common one, the solid undergoes large elastic deformations, for instance in crops or aquatic vegetation. In the second type, the system breaks before any significant deformation can occur; this will be referred to as brittle behavior in the following. The former has been abundantly studied, a key result being that of load reduction by elastic reconfiguration \cite{vogel_1989,gosselin_2010}. The latter has already been described in trees or corals \cite{koehl_1984,niklas_1999}, but to the best of our knowledge the effect of branching has never been studied theoretically. Therefore, we shall focus hereafter on brittle branched slender systems, which are ubiquitous in nature: trees \cite{mcmahon_1975}, bushes, algae \cite{koehl_1984}, corals \cite{madin_2005} and corallines \cite{martone_2008b}, to list a few. In the following we refer mainly to trees under wind loading, with the understanding that these results are also applicable to a large variety of other biological systems under fluid-loading. 

\begin{figure*} [tb]
	\centering
		\includegraphics[width=0.85\textwidth]{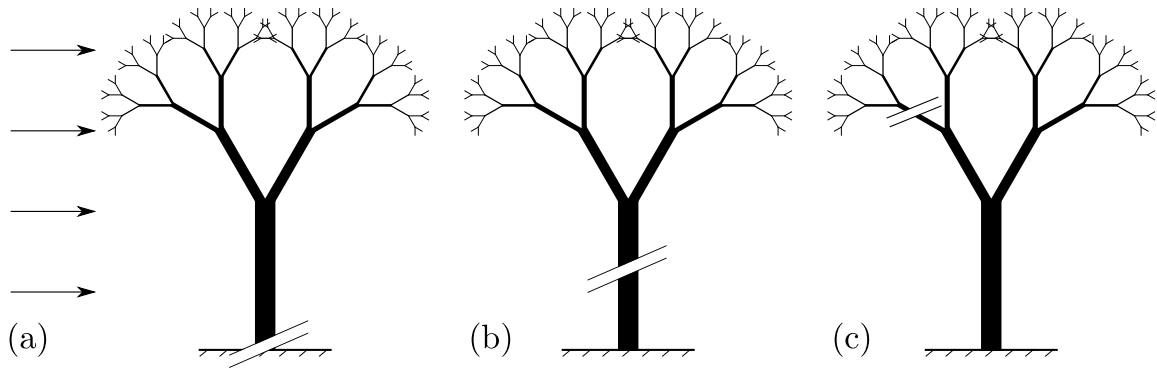}
	\caption{Schematic view of breakage process in a branched brittle system under flow. (a) Base breakage, (b) Trunk breakage, (c) Branch breakage.}
	\label{fig:1}
\end{figure*}

For a brittle branched system attached to a support, breakage under flow may occur in three distinct types: (i) base breakage, Fig.~\ref{fig:1}a, when the attachment to the ground is broken, as in uprooting, (ii) trunk breakage, Fig.~\ref{fig:1}b, when the main element is broken, and (iii) branch breakage, Fig.~\ref{fig:1}c, when an upper element breaks, as in flow-induced pruning.

In fact, the distinction between trunk and branch breakage has a biological relevance, since breakage of the trunk is likely to be fatal, while re-growth is often possible after branch breakage. Moreover branch breakage does reduce loads on the trunk and the attachment, as in elastic reconfiguration, thereby delaying their breakage \cite{koehl_1984,niklas_2000}. 
Finally, branch breakage can also be part of the asexual reproduction process by propagation. This is observed in terrestrial plants such as willows and poplars \cite{beismann_2000}, and in stony corals such as \textit{Acropora Cervicornis} or \textit{Acropora Palmata} \cite{tunnicliffe_1981,highsmith_1982}.

Breakage is the consequence of an unacceptable stress level; it is therefore directly related to the stress state in the structure \cite{niklas_2000,gardiner_forest_2000}. In particular, the issue of whether the stress level is uniform or not in the tree is crucial, as breakage is expected to occur at the point of maximal stress. For instance, Niklas and Spatz \cite{niklas_2000} showed that in a cherry tree the stress level varies by one to two orders of magnitude within the tree and has a local maximum in the branches. On the other hand, Bejan et al.~showed that the flow-induced stress is uniform for a tapered trunk when the taper is linear \cite{bejan_2008}. In fact the stem taper is an important parameter regarding the stress distribution; see the discussion in \cite{larjavaara_2010}.

\begin{figure*} [tb]
	\centering
		\includegraphics[width=0.85\textwidth]{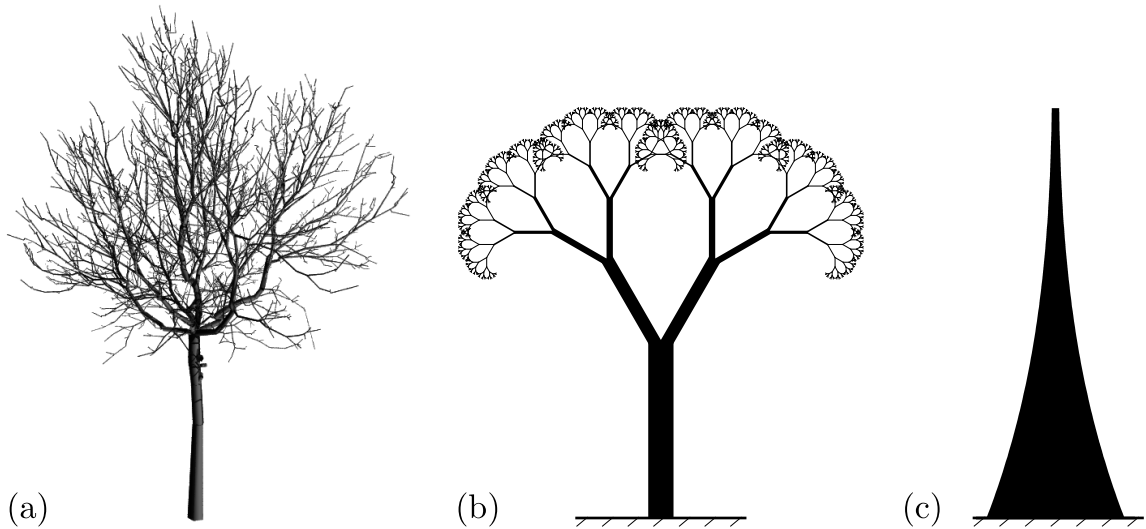}
	\caption{Geometries of the models used in the paper: (a)~Section \ref{sec:FlowInducedPruningOfAWalnutTree}: Walnut tree, as in \cite{sinoquet_1997}; (b)~Section \ref{sec:TheIdealTreeModel}: Idealized branched system, as in \cite{rodriguez_2008}; (c)~Section \ref{sec:TheSlenderConeModel}: Tapered beam, as in \cite{mcmahon_1975,bejan_2008}.}
	\label{fig:2}
\end{figure*}

Several questions remain however regarding the flow-induced breakage of tree-like structures: (i) what are the effects of the geometrical and physical parameters on the occurrence of branch breakage? (ii) How do the breaking events occur successively as the flow is increased? (iii) Assuming that branch breakage is favorable in biological terms, is it compatible with other constraints on the geometry? The aim of this paper is to address these questions, using simple numerical and analytical models for the mechanical behavior of slender and brittle structures. The modeling assumptions and framework used throughout the paper are first presented in Section \ref{sec:MechanicalModelAndParameters}. In Section \ref{sec:FlowInducedPruningOfAWalnutTree}, we compute the stress distribution and successive breaking events in a complex tree, using the geometry of an actual walnut tree. Using an idealized branched system, we derive conditions for branch breakage in Section \ref{sec:TheIdealTreeModel}. These are further analyzed for a tapered beam, here referred to as the slender cone model, in Section \ref{sec:TheSlenderConeModel}. The corresponding three geometries are sketched in Fig.~\ref{fig:2}. Finally a general discussion and conclusion are given in Section \ref{sec:DiscussionAndConclusions}.

\section{Mechanical model and parameters}
\label{sec:MechanicalModelAndParameters}

Throughout the paper, we consider a cross flow over the entire structure, uniform, as the dependence of the stress on the wind velocity profile was shown to be small \cite{niklas_2000}. Also, only static loads are taken into account, and the corresponding fluid force magnitude $f$ per unit length reads
\begin{equation}
	f = \frac{1}{2} \rho C_D D U^2,
	\label{eq:flin}
\end{equation}
where $U$ is the free stream velocity, $\rho$ its density, $D$ the local branch diameter and $C_D$ the drag coefficient \cite{delangre_2008,madin_2006}. 
The direction is assumed to be that of the flow velocity. The fluid load is here computed on a leafless branch, and the influence of leaves will be discussed in Section \ref{sec:DiscussionAndConclusions}.

This load is applied on the whole branched system, which is held by a perfect clamping at the base. Because of the high slenderness of the system, we use a standard linear beam theory to derive the stress state, essentially the bending moment $M$. The maximum stress in the cross-section resulting from this bending moment is the skin stress, defined as $\Sigma = 32M/\pi D^3$ \cite{niklas_1992,gere_1990}.

The brittle behavior is introduced as follows: (i) the deformations are assumed to be negligible, so the stress state is computed on the initial configuration, without elastic reconfiguration, (ii) when increasing the flow velocity $U$, breakage occurs when and where the local skin stress $\Sigma$ reaches a critical value, $\Sigma_c$. Then, the broken branch is removed, and this results in a new flow-induced stress state. Flow velocity may then be further increased until a new breaking event occurs. 

Throughout the paper, the relevant dimensionless number to scale the fluid-loading $\rho C_D U^2$ with respect to the critical stress $\Sigma_c$ is the Cauchy number, defined as
\begin{equation}
	C_Y = \frac{\rho C_D U^2}{\Sigma_c} G ,
	\label{eq:cy}
\end{equation}
where $G$ is a geometrical factor introduced for comparison purpose and defined such that $\Sigma = \Sigma_c$ at the base of the intact structure when $C_Y=1$. Note that this Cauchy number is similar in principle but differs from that used in the analysis of flow-induced elastic deformation, namely $C_Y = \rho C_D U^2/E$ \cite{delangre_2008,gosselin_2010}; the critical stress $\Sigma _c$ simply replaces here the Young modulus $E$.

The non-dimensional stress is defined as $\sigma = \Sigma/\Sigma_c$ and the non-\-dimen\-sional bending moment as $m=M/M_c$, with $M_c=\Sigma_c \pi D_B^3/32$, $D_B$ being the base diameter \cite{niklas_1992}. This latter scaling is chosen so that failure occurs at the base of the trunk when $m=1$. The non-dimensional vertical coordinate $z$ is defined using $H$, the height of the structure, as a reference length scale.

\section{Flow-induced pruning of a walnut tree}
\label{sec:FlowInducedPruningOfAWalnutTree}

\begin{figure*} [tb]
	\centering
		\includegraphics[width=0.85\textwidth]{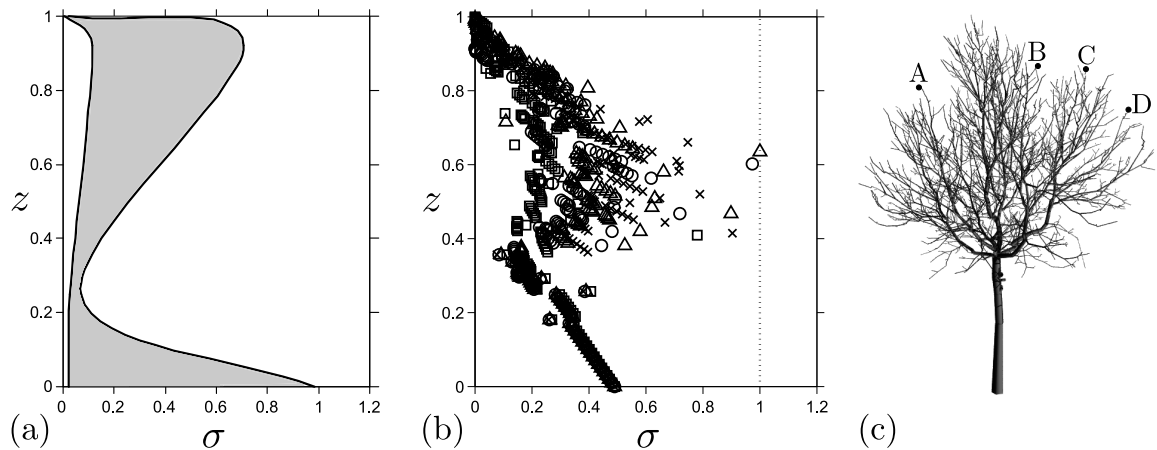}
	\caption{Non-dimensional stress profile $\sigma$ in a tree under cross-flow. (a) Schematic view of the stress profiles given by Niklas and Spatz \cite{niklas_2000} for cherry trees, showing a local maximum near the top. (b)~Computed stress profiles along four branching paths, A ($\times$), B ($\square$), C ($\triangle$) and D ({\large $\circ$}) in the digitized tree geometry shown in~(c).}
	\label{fig:3}
\end{figure*}

The geometry of the branched system is expected to have a large influence on the stress state and thus on the location and timing of breaking events. We therefore first apply the procedure described above using the digitized geometry of an actual 20-yr-old walnut tree (\textit{Juglans Regia L.}) described in \cite{sinoquet_1997} (Fig.~\ref{fig:2}a). This tree is 7.9 m high, 18 cm in diameter at breast height (dbh), and has a sympodial branching pattern \cite{barthelemy_2007} and about eight orders of branching. The stress state under flow is computed using a standard finite element software (CASTEM v. 3M \cite{verpeaux_1988}), and is presented in Fig.~\ref{fig:3}b for four different branching paths.

\begin{figure} [tb]
	\centering
		\includegraphics[width=0.4\textwidth]{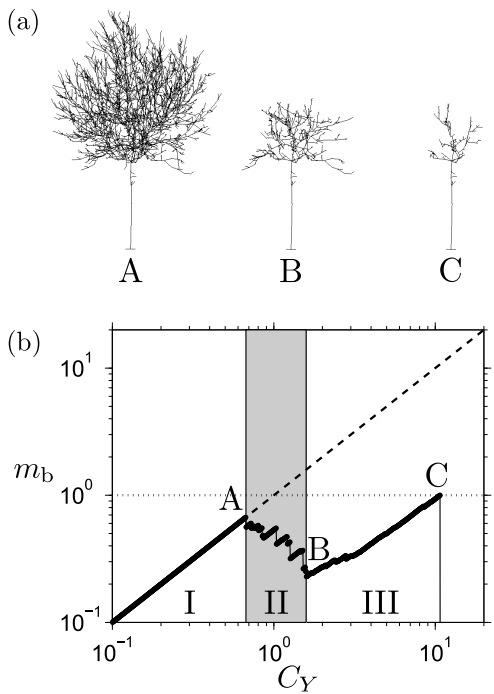}
	\caption{Computed sequence of branch breakage in the walnut tree: (a)~A: initial tree for $C_Y \leq 0.67$; B: after breakage in large branches, $C_Y = 1.7$; C: just before trunk breakage, $C_Y = 10.7$. (b)~Corresponding evolution of the bending moment at the base of the tree $m_\text{b}$, in three distinct ranges. The dashed line shows the moment that would exist without breakage. The dotted line shows the critical value $m_\text{b}$ that causes trunk breakage.}
	\label{fig:4}
\end{figure}

We observe that the stress level is not uniform but shows a maximum located in the branches, which is consistent with the results of Niklas and Spatz \cite{niklas_2000} which are sketched in Fig.~\ref{fig:3}a. Note that since $\sigma$ varies linearly with the fluid-loading $C_Y$, one needs only to focus on the critical situation where $\sigma=1$ is first reached in the structure.
In this tree, the criterion for breakage is satisfied first in a branch and not in the trunk. This corresponds to the mechanism of branch breakage, as defined in Section \ref{sec:Introduction}. If the fluid-loading is further increased after removal of the broken parts, successive breaking events are observed, in a flow-induced pruning sequence: Fig.~\ref{fig:4}a shows three states of the tree at increasing Cauchy number with branches progressively removed as they break off.

During the sequence of breakage, the bending moment at the base of the tree, $m_\text{b}$, evolves significantly with the Cauchy number, Fig.~\ref{fig:4}b. Up to the first breakage, the moment is proportional to the fluid-loading $C_Y$ (zone I in Fig.~\ref{fig:4}b). Then, in a small range of load increase (zone II), all large branches are broken at an intermediate level, resulting in a significant decrease of the bending moment. Breakage then continues but to a much smaller extent (zone III), while the moment increases almost linearly up to the value $m_\text{b}=1$ when the trunk breaks. Note that the benefit of this sequence of breaking events is that the critical value of the base moment $m_\text{b}=1$ is reached only at $C_Y \simeq 10$ instead of $C_Y=1$ if there was no branch breakage. This corresponds to more than a factor of 3 on the acceptable fluid velocity.
For instance, for a critical stress $\Sigma_c = 40$ MPa, which is the order of magnitude of maximum acceptable bending stresses measured in trees \cite{beismann_2000, lundstrom_2008}, the maximum sustainable fluid velocity before trunk breakage is increased from $U \simeq 30 \text{ m.s}^{-1}$ without branch breakage to $U \simeq 100 \text{ m.s}^{-1}$ with branch breakage.

To summarize, this set of computations clearly shows that branch breakage can occur prior to trunk breakage, and that the sequence of flow-induced pruning results in a significant reduction in the load applied on the base of the tree, or equivalently, an increase in the sustainable fluid velocity. To further analyze this process, we turn to a simple model in the next section.

\section{The ideal tree model}
\label{sec:TheIdealTreeModel}

\subsection{Infinite branched tree}
\label{sec:InfiniteBranchedTree}

\begin{figure} [tb]
	\centering
		\includegraphics[width=0.46\textwidth]{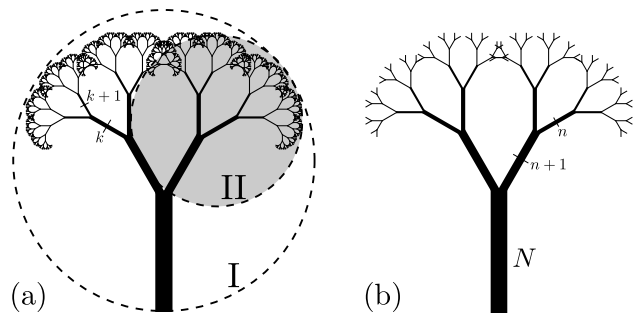}
	\caption{Idealized branched system. (a)~Infinite iterated tree. The sub-tree II is equivalent to the whole tree I but for a change of scales. (b)~Finite iterated tree and corresponding notations.}
	\label{fig:5}
\end{figure}

To establish the relation between the parameters of the system and the flow-induced pruning process, we simplify the problem to its essential elements: the branched geometry and the slenderness of branches; we disregard here the effect of branch orientation relative to the flow. Similarly to \cite{rodriguez_2008}, we consider first an infinitely iterated sympodial tree made of cylindrical branches (Fig.~\ref{fig:5}). Two parameters only are needed to describe this ideal tree: (i) the branching ratio $\lambda$, giving the reduction of diameter through branching, and (ii) the slenderness exponent $\beta$, giving the relationship between length and diameter in branch segments of the tree, so that
\begin{equation}
	\lambda = \left(\frac{D_{k+1}}{D_k}\right)^2 ,\quad \frac{D_{k+1}}{D_k}=\left(\frac{L_{k+1}}{L_k}\right)^\beta ,
\end{equation}
where $D_k$ and $L_k$ are the diameter and length of a branch segment of order $k$, see Fig.~\ref{fig:5}a \cite{rodriguez_2008}. Typical values of these parameters are $\lambda<1$ and $1<\beta<2$. Note that the number of branches emerging from a branching point is typically equal to $1/\lambda$ \cite{lindenmayer_1996}.

We use now a scaling argument similar to that of \cite{rodriguez_2008} for the dynamics of trees. On the ideal infinitely branched system of Fig.~\ref{fig:5}a, we can compare the stress level in branch $k=1$ (the trunk) and in branch $k=2$. The sub-tree labeled II in Fig.~\ref{fig:5}a is identical to the full tree, I, but for a change in length and diameter scales. All diameters (resp. lengths) in II are reduced by a factor $\lambda^{1/2}$ (resp. $\lambda^{1/2\beta}$). Let $\Sigma_1$ be the maximum skin stress in the trunk ($k=1$) under a given fluid-loading $U$, and $\Sigma_2$ the maximum skin stress in the branch $k=2$.
The relations between the flow velocity and $\Sigma_1$ or $\Sigma_2$ are identical, but for the change of diameter and length scales. The dependence of the stress on diameter and length is the following: (i) $\Sigma$ varies as $M/D^3$, where $M$ is the bending moment, (ii) $M$ varies as $fL^2$, where $f$ is the norm of the local fluid force, Eq.~\eqref{eq:flin}, (iii) $f$ varies as $\rho U^2 D$. Hence $\Sigma$ varies as $\rho U^2(L/D)^2$. We therefore may state that
\begin{equation}
	\frac{\Sigma_2}{\Sigma_1} = \left(\frac{L_2}{D_2}\right)^2 \left(\frac{D_1}{L_1}\right)^2 = \lambda^\frac{1-\beta}{\beta}.
\end{equation}
Since $\lambda<1$, the condition for the stress to be higher in branches than in the trunk becomes
\begin{equation}
	\beta>1.
\end{equation}

Here the only parameter controlling the possibility of branch breakage is the slenderness exponent, a classical parameter in the allometry of trees. As $\beta$ is typically greater than 1 for trees, branch breakage is expected to occur. This simplistic approach now deserves to be improved, as the assumption of an infinite number of branching levels is very strong, and may not be compatible with the constraint that the tree area has to be finite.

\subsection{Finite branched tree}
\label{sec:FiniteBranchedTree}

Let us consider now the same idealized tree, but with a finite number of branching iterations (Fig.~\ref{fig:5}b). This structure has $N$ levels, which are labeled in this section from the top to the bottom. Note that $n=N-k+1$, where $n$ is the label of the previous section from the base of the tree. The trunk corresponds now to the last level, $N$. At each level $n$, we define the branch diameter $D_n$ and length $L_n$, which can be expressed as a function of the trunk diameter and length $D_N$ and $L_N$ as
\begin{equation}
	D_n = \lambda^{\frac{N-n}{2}} D_N,\quad L_n = \lambda^{\frac{N-n}{2\beta}} L_N.
	\label{eq:dlrec}
\end{equation}

By a simple integration of the fluid force on the branches, the moment at the base of a branch of order $n$ may be derived, as well as the corresponding skin stress, which is obtained in non-dimensional form as
\begin{equation}
	\sigma_n = C_Y \lambda^{\frac{1-\beta }{\beta }N} \left( A \lambda ^{\frac{\beta -1}{\beta }n} + B \lambda ^{\frac{n}{2}} + C \lambda ^{\frac{\beta -1}{2 \beta }n} \right),
	\label{eq:sigtree}
\end{equation}
where the Cauchy number $C_Y$ is defined as
\begin{equation}
	C_Y = \left[\frac{8}{\pi} \left(\frac{L_N}{D_N}\right)^2\right]\frac{\rho C_D U^2}{\Sigma_c},
	\label{eq:cytree}
\end{equation}
and $A$, $B$ and $C$ are functions of $\beta$ and $\lambda$ only. The detailed derivation of Eq.~\eqref{eq:sigtree} as well as the expression of $A$, $B$ and $C$ can be found in \ref{sec:StressCalculationInBranchedSystem}.

\begin{figure} [tb]
	\centering
		\includegraphics[width=0.35\textwidth]{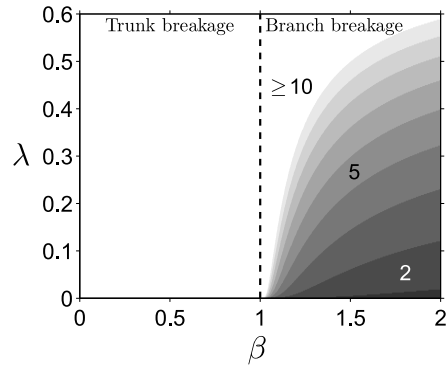}
	\caption{Location of the maximum of stress under cross-flow in an idealized tree model, as a function of the slenderness exponent $\beta$ and the branching parameter $\lambda$. The location is given in the form of the number of branching levels counted from the top of the tree, Fig.~\ref{fig:5}b. For $\beta\leq 1$, the breakage is directly at the base of trunk.}
	\label{fig:6}
\end{figure}

A systematic numerical exploration of the $(\lambda, \beta$) parameter space shows that when $\beta<1$ the stress always increases from top to bottom. Conversely, for $\beta>1$, the stress reaches a maximum at branch level $n_c$ and then decreases from top to bottom, provided that $N>n_c$, where $n_c$ depends on $\lambda$ and $\beta$. This dependence is given in Fig.~\ref{fig:6}. This analysis with a finite tree model gives a criterion consistent with that of the infinite tree model, namely $\beta>1$. Moreover, the other parameter, $\lambda$, is found to affect only the location of possible breakage. This suggests that branching is not a key factor in the occurrence of branch or trunk breakage. In the next section we explore a simpler model of the slenderness effect.

\section{The slender cone model}
\label{sec:TheSlenderConeModel}

\subsection{Flow-induced stress}
\label{sec:FlowInducedStress}

The simplest model that allows one to take into account a relation between diameters and lengths through a slenderness exponent is a cone. This formulation is related to MacMahon and Kronauer's equivalent geometry of a tree, a tapered beam with a rectangular cross-section of dimensions varying as power laws of height \cite{mcmahon_1975,mcmahon_1976}.

\begin{figure} [tb]
	\centering
		\includegraphics[width=0.35\textwidth]{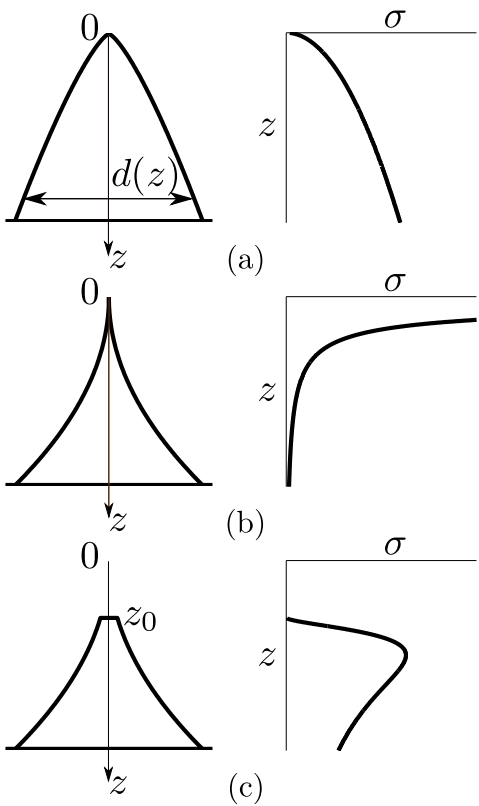}
	\caption{The slender cone model: geometry and stress profile under uniform cross flow. (a)~cone with $\beta<1$ (here 0.75), showing a maximum of stress at the base; (b)~cone with $\beta>1$ (here 2), showing a maximum at the top; (c)~cone truncated arbitrarily at $z_0=0.3$ showing a local maximum.}
	\label{fig:7}
\end{figure}

The geometry considered here is a slender cone with a circular cross-section, Fig.~\ref{fig:7}a, and we follow the same mechanical approach as for the previous geometries. Let $H$ be the cone height, $d_H = D_H/H$ the dimensionless diameter at the base and  $z$ the vertical coordinate which is orientated downwards in this section. The cone dimensionless diameter is given by
\begin{equation}
	d(z)=d_H z^\beta.
	\label{eq:dz}
\end{equation}

Using the same formulation as in the previous section, the stress state along the cone is obtained as
\begin{equation}
	\sigma(z) = C_Y z^{2(1-\beta)},
	\label{eq:sig_inf}
\end{equation}
where the Cauchy number is defined here as
\begin{equation}
	C_Y = \left[\frac{16}{(1+\beta)(2+\beta)\pi d_H^2}\right] \frac{\rho C_D U^2}{\Sigma_c}.
	\label{eq:cycone}
\end{equation}
From Eq.~\eqref{eq:sig_inf}, we readily observe that: (i) for $\beta=1$, the constant stress case of Bejan et al.~\cite{bejan_2008} is found; (ii) for $\beta<1$ the stress increases with $z$ and is therefore maximum at the base, Fig.~\ref{fig:7}a; (iii) for $\beta>1$ the stress decreases with $z$, and the maximum, discussed further, is not at the base, Fig.~\ref{fig:7}b-c. These results are consistent with the condition for branch breakage in the previous section.

To avoid the singular case of infinite stress at $z=0$ for $\beta>1$, we use a cone truncated at $z=z_0$, Fig.~\ref{fig:7}c. The truncation $z_0$ corresponds to the first breakage occurring as soon as $U \neq 0$, and its value is chosen arbitrarily. The corresponding stress state is then
\begin{equation}
	\frac{\sigma(z)}{C_Y} = z^{2(1-\beta)} -(2+ \beta)z_0^{1+\beta}z^{1-3\beta} + (1+\beta)z_0^{2+\beta}z^{-3\beta},
	\label{eq:sig_trunc}
\end{equation}
which reduces to Eq.~\eqref{eq:sig_inf} when $z_0=0$. The detailed derivation of this equation is given in \ref{sec:StressCalculationInSlenderCone}. For $\beta>1$, the stress shows a maximum before decreasing downwards, as illustrated in Fig.~\ref{fig:7}c. The limit case $z_0 = 0$ is in fact equivalent, in the ideal tree model of Section \ref{sec:TheIdealTreeModel}, to the limit as $N$ goes towards infinity, which would lead to a vanishing diameter at the tip. There is therefore an analogy between the cone truncation and the ideal tree with a finite number of branching levels.

\subsection{Sequence of breaking events}
\label{sec:SequenceOfBreakingEvents}

Considering now the generic case of the truncated cone, Fig.~\ref{fig:7}c, we analyze the sequence of breaking events resulting from an increasing fluid-loading $C_Y$. The stress $\sigma$ increases linearly with $C_Y$ up to the point where its maximum value reaches the limit of breakage, $\sigma=1$. This defines the first breaking event at $C_Y = C_Y^1$ occurring at $z=z_1$. It results in a new truncated cone, and the process is repeated as $C_Y$ is further increased. Eventually, when the cone becomes truncated close to the base, the maximum stress may be reached at the base itself, resulting finally in base breakage.

\begin{figure} [tb]
	\centering
	\includegraphics[width=0.4\textwidth]{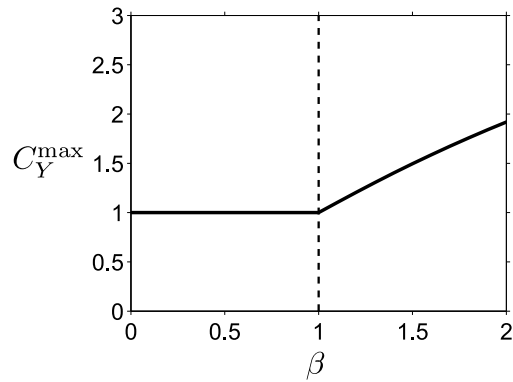}
	\caption{Maximum fluid load that the cone can support as a function of the slenderness exponent. Note that for $\beta>1$ the curve is the lower bound of all possible evolutions.}
	\label{fig:8}
\end{figure}

This sequence of breaking events may be analyzed in terms of the maximum fluid-loading $C_Y^\text{max}$ that the cone can support before breaking at the base. As illustrated in Fig.~\ref{fig:8}, this is strongly dependent on $\beta$. When $\beta<1$, the first breaking event is at the base so that $C_Y^\text{max}=1$. Conversely when $\beta>1$, breaking occurs progressively as $C_Y$ is increased, and the base breakage is delayed, $C_Y^\text{max}>1$. The precise value of $C_Y$ where the base breaks depends on the initial truncation $z_0$, but is always higher than a lower bound that can be computed from Eq.~\eqref{eq:sig_trunc}, which is shown in Fig.~\ref{fig:8}. We observe a significant increase of the ability of the system to sustain fluid-loading when $\beta>1$.

\begin{figure} [tb]
	\centering
	\includegraphics[width=0.4\textwidth]{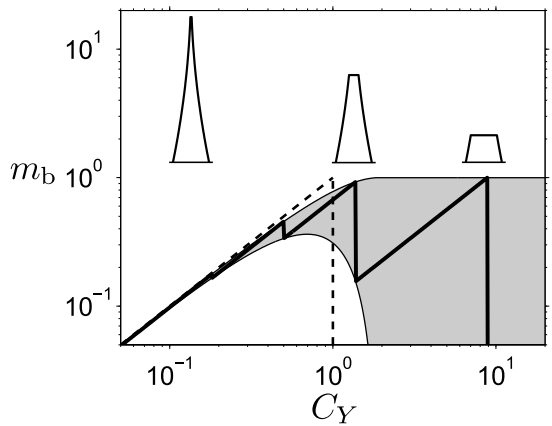}
	\caption{Moment at the base of the cone as the fluid-loading is increased. (-~-~-) direct base breakage occurring when $\beta<1$; (---) progressive breaking for $\beta>1$ (here $\beta=2$). The shaded region shows all possible values depending on the initial truncation $z_0$. The cone state is shown for three values of $C_Y$.}
\label{fig:9}
\end{figure}

In terms of base moment, the sequence of breaking events can be easily computed, Fig.~\ref{fig:9}. For $\beta<1$ the base moment increases linearly with $C_Y$ until base breakage occurs, $m_\text{b}=1$ for $C_Y=1$. For $\beta>1$ the sequence of breaking events results in sudden drops in base moment followed by linear increase up to the next breaking, as illustrated in Fig.~\ref{fig:9}. Since the sequence of breaking events is a discrete process that depends on the initial truncation $z_0$, there exists, for a given Cauchy number $C_Y$, a wide range of acceptable cone heights and thereby a wide range of corresponding base moments.
In practice, for all possible values of $z_0$, the evolution of $m_\text{b}$ remains bounded between its values for the shortest and highest cone that can exist at each Cauchy number. This is represented by the shaded region in Fig.~\ref{fig:9}.

These results show that the simple cone model contains the key elements to understand the effect of geometry on (i) the stress profile, (ii) the sequence of breaking events and (iii) the consequences on the evolution of base load when the fluid velocity is increased. Here again, the essential criterion concerns the slenderness exponent $\beta$.

\section{Discussion and conclusions}
\label{sec:DiscussionAndConclusions}

Starting from the case of a full walnut tree geometry, we have used models of increasing simplicity. This allowed us to point out the role of various parameters on the process of breakage under fluid-loading.
The first issue that had to be addressed was that of the flow-induced stress distribution. As noted by other authors, the stress is not necessarily maximum at the base \cite{niklas_2000,bejan_2008}. In fact in the walnut tree of Section \ref{sec:FlowInducedPruningOfAWalnutTree}, the stress has a local maximum at about mid height. Using the ideal tree model in Section \ref{sec:TheIdealTreeModel}, we have shown that the existence of this maximum is related to the value of the slenderness exponent, $\beta$, being larger than one: in fact this allometry parameter is about 1.37 for this particular walnut tree \cite{rodriguez_2008}. Following Bejan et al.~\cite{bejan_2008}, we recover the critical value of $\beta=1$ in the simplest model, that of a cone in Section \ref{sec:TheSlenderConeModel}.

Actually, some refinement is needed here to understand the precise location of the maximum of stress. We have shown in Section \ref{sec:TheIdealTreeModel} that the location of this maximum was also dependent on the branching parameter $\lambda$, in the form of the parameter $n_c$, which is the number of branching levels from the top to this maximum point. For our walnut tree, where $\lambda \simeq 0.25$, we obtain $n_c=6$ using Fig.~\ref{fig:6}. This is smaller than the total number of branching levels in the walnut tree which is about 8 \cite{sinoquet_1997}. A local maximum of stress is therefore expected in the branches, and is actually observed in Fig.~\ref{fig:3}.

The second issue was that of the sequence of breaking events occurring when the fluid-loading $C_Y$ is increased. Using a brittle fracture model for the walnut tree in Section \ref{sec:FlowInducedPruningOfAWalnutTree}, we have shown that most large branches broke in a short range of flow velocity, and that breakage of the trunk occurred much later. The large size of broken branches can be explained by the value of $n_c=6$ found above. All large branches do not break exactly at the same value of the Cauchy number. This is due among other reasons to some variability in the allometry parameters $\lambda$ and $\beta$ within the tree. Once all large branches are broken, the remaining tree shape, C in Fig.~\ref{fig:4}a, does not have enough branching levels to have a local maximum, and the next breaking event occurs at the base of the trunk. 
Note that the process of branch breakage in the walnut tree allowed the tree to have a much larger acceptable Cauchy number before breakage of the trunk. This can also be analyzed using the cone model as in Section \ref{sec:TheSlenderConeModel}, where the critical Cauchy number for base breakage is clearly dependent on $\beta$ (Fig.~\ref{fig:8}).

The third issue was that of the evolution of the load at the base of the tree. For the walnut tree, Fig.~\ref{fig:4}b, the sequence of successive breakage of the large branches results in a significant decrease of the drag-induced moment at the base. This can be understood using the cone model, where the sequence of breaking event and corresponding drops of base moment can be tracked, Fig.~\ref{fig:9}. We may therefore state that the essential characteristics of branch breakage and corresponding load evolution in the walnut tree can be understood using our simple ideal tree model and cone model.

The analytical results of Sections \ref{sec:TheIdealTreeModel} and \ref{sec:TheSlenderConeModel} were obtained considering that all parameters have self-similar variations. However, this was not the case for the walnut tree computations of Section \ref{sec:FlowInducedPruningOfAWalnutTree}, which suggests that the behaviors pointed out in this study can be generalized to structures that do not necessarily have self-similar variations of their parameters.
Moreover, the ideal tree and cone models can be easily extended to incorporate other features of the problem, such as a dependence of all parameters with $z$: the flow velocity $U$, the material properties through the critical parameter $\Sigma_c$, and even the drag coefficient $C_D$, which allows one to take easily into account the additional drag of leaves.
Preliminary results, not shown here for the sake of brevity, showed that the criterion for branch breakage takes the same form, but involves both $\beta$ and the corresponding parameter related to the additional $z$-dependence. Taking into account a significant elastic deformation before load fracture, or incorporating dynamical effects, would be much more complex.

\begin{table*}
	\tiny
	\begin{center}
	\begin{tabular}{ccccccc}
  	\hline
  	\multirow{2}{*}{Ref.} & \multirow{2}{*}{Tree} & Slenderness & Branching & Total orders of & Predicted branch & Predicted \rule[0pt]{0pt}{13pt} \\ 
  	& & exponent $\beta$ & parameter $\lambda$ & branching $N$ & breakage level $n_c$ & breakage type \rule[-9pt]{0pt}{13pt}\\ \hline 
		\multirow{2}{*}{\cite{sinoquet_1997,rodriguez_2008}} & Walnut Tree &  \multirow{2}{*}{1.37} & \multirow{2}{*}{0.25} & \multirow{2}{*}{$>8$} & \multirow{2}{*}{6} & \multirow{2}{*}{Branch} \rule[0pt]{0pt}{13pt} \\
		& \textit{Juglans Regia L.} &&&&&\\
		\multirow{2}{*}{\cite{mcmahon_1976}} & Red Oak & \multirow{2}{*}{1.51} & \multirow{2}{*}{0.41} & \multirow{2}{*}{$>6$} & \multirow{2}{*}{7} & Branch or \rule[0pt]{0pt}{10pt} \\
		& \textit{Quercus Rubra} &&&&& Trunk \\
		\multirow{2}{*}{- -}& White Oak 1 & \multirow{2}{*}{1.41} & \multirow{2}{*}{0.28} & \multirow{2}{*}{$>6$} & \multirow{2}{*}{6} & \multirow{2}{*}{Branch} \rule[0pt]{0pt}{10pt} \\ 
		& \textit{Quercus Alba} &&&&&\\
		\multirow{2}{*}{- -}& White Oak 2 & \multirow{2}{*}{1.66} & \multirow{2}{*}{0.29} & \multirow{2}{*}{$>6$} & \multirow{2}{*}{5} & \multirow{2}{*}{Branch} \rule[0pt]{0pt}{10pt} \\
		& \textit{Quercus Alba} &&&&&\\
		\multirow{2}{*}{- -}& Poplar Tree & 1.5 & \multirow{2}{*}{0.29} & \multirow{2}{*}{$>6$} & \multirow{2}{*}{5} & \multirow{2}{*}{Branch} \rule[0pt]{0pt}{10pt} \\
		& \textit{Populus Tremoloides} & (estimated) &&&&\\
		\multirow{2}{*}{- -}& Pin Cherry & \multirow{2}{*}{1.5} & \multirow{2}{*}{0.24} & \multirow{2}{*}{$>4$} & \multirow{2}{*}{5} & Branch or \rule[0pt]{0pt}{10pt} \\
		& \textit{Prunus Pensylvanica} &&&&& Trunk \\
		\multirow{2}{*}{- -}& White Pine & \multirow{2}{*}{1.37} & \multirow{2}{*}{0.24} & \multirow{2}{*}{$>5$} & \multirow{2}{*}{5} & \multirow{2}{*}{Branch} \rule[0pt]{0pt}{10pt} \\
		& \textit{Pinus Strobus} &&&&&\rule[-6pt]{0pt}{10pt}\\\hline
	\end{tabular}
	\caption{Predicted breakage type using the results of Section \ref{sec:TheIdealTreeModel}. Branch breakage is predicted when $n_c \leq N$.}
	\label{tab:1}
	\end{center}
\end{table*}

Considering the simplicity of the criterion that we have found for branch breakage, we can test whether it is generally satisfied. MacMahon and Kronauer \cite{mcmahon_1976} have noted that $\beta$ is usually larger than 1 and typically around 1.5, while $\lambda$ is typically close to 0.25. This leads to a maximum stress located at a branching level $n_c=5$ counting from top down. This is clearly in the branches as trees generally have more than 5 orders of branching. We may therefore state that branch breakage can be expected in most sympodial trees. This is illustrated in Table~\ref{tab:1}, where the values of parameters are given for several trees.

Clearly the possibility of branch breakage is favorable in terms of survival of an individual tree in the face of extreme fluid-loading. It may also be favorable in terms of tree development by removing the less vigorous branches. The question then arises as to whether this implies new constraints on the geometry of the tree. It appears from our results that the constraint $\beta>1$ is not incompatible with other constraints such as the optimal resistance to buckling under gravity, which requires $\beta=3/2$ \cite{mcmahon_1975}. The same result was obtained considering the wind effect on trees but for an overcrowded tree canopy \cite{larjavaara_2010}. Similarly $\beta>1$ is compatible with a constraint for optimal dissipation \cite{rodriguez_2008,theckes_2011}, that modal frequencies have a ratio of less than two, requiring that $\beta>1$ for $\lambda = 0.25$.

The particular case of branched corals \cite{madin_2005,tunnicliffe_1981,highsmith_1982} is somewhat different. The segments are similar in length and diameter, so that $\lambda \simeq 1$ and $\beta \simeq 1$ in our variables, but with a number of branches emerging from one branching not equal to $1/\lambda$. An analysis similar to that of Section~\ref{sec:TheIdealTreeModel} shows that breakage is expected at the bottom. This is the case in most isolated corals.

More generally we may place these results in the overall context of reconfiguration, as introduced by Vogel \cite{vogel_1989}. This originally referred to the reduction of loading made possible by elastic deformation. For a plant, it is a crucial mechanism to survive heavy fluid-loading. But plant tissues are not all very elastic, and plant parts are not all very flexible. Our results on the role of branch breakage in reducing loading show that, in parallel with elastic reconfiguration, there exists a mechanism of brittle reconfiguration. There are therefore two distinct strategies to overcome critical events. The first is evidently reversible in the short term by elasticity. The second is also reversible by re-growth, but only in the long term. Thus flow-induced pruning is possibly a widespread mechanism in plants or benthic organisms that support heavy loading by the surrounding fluid environment.

\section*{Acknowledgements}
The authors gratefully acknowledge the help of Chris Bertram, from the University of Sydney, for stimulating discussions and useful corrections on the manuscript.
The authors also acknowledge the interesting comments and useful suggestions from the anonymous reviewers. The first author was funded by the PhD scholarship program ``AMX'' at Ecole Polytechnique.

\appendix
\section{Stress derivation in finite branched tree model}
\label{sec:StressCalculationInBranchedSystem}

In order to compute the stress along the finite ideal tree, we introduce $f_n$ the fluid force per unit length at level $n$, $f_n = \frac{1}{2}\rho C_D U^{2} D_n$, with the same notations as Eq.~\eqref{eq:flin}. At each level $n$, we consider two force components: (i) the shear force $\tau_n$ in the flow direction and (ii) the bending moment $M_n$ in the direction normal to the flow. Due to the free condition at the top, $\tau_0 = 0$ and $M_0 = 0$, and for $n \geq 1$
\begin{eqnarray}
	\label{eq:Tau}
	\tau_n & = & f_n L_n + p \tau_{n-1},\\
  \label{eq:Mn}
	M_n & = & \frac{1}{2} f_n L_n^2+p \left(M_{n-1}+L_n \tau_{n-1}\right),
\end{eqnarray}
where $p$ is the number of branches emerging from one at a branching point $\left(p=1/\lambda\right)$.
The non-dimensional stress $\sigma_n$ at level $n$ reads
\begin{equation}
	\sigma_n = \frac{32M_n}{\pi \Sigma_c D_n^3}
\end{equation}
By integration of Eqs.~\eqref{eq:Tau} and \eqref{eq:Mn}, the stress at each level can be obtained,
\begin{equation}
	\sigma_n = C_Y \lambda^{\frac{1-\beta }{\beta }N} \left( A \lambda ^{\frac{\beta -1}{\beta }n} + B \lambda ^{\frac{n}{2}} + C \lambda ^{\frac{\beta -1}{2 \beta }n} \right),
\end{equation}
with
\begin{equation}
	C_Y = \left[\frac{8}{\pi} \left(\frac{L_N}{D_N}\right)^2\right]\frac{\rho C_D U^2}{\Sigma_c},
\end{equation}
and
\begin{eqnarray}
	A &=&  \frac{ \lambda ^{\frac{1-\beta}{2 \beta }}+1}{\left(\lambda ^{\frac{1-\beta}{2 \beta }}-1\right) \left(\lambda ^{\frac{2-\beta}{2 \beta }}-1\right)},\\
	B &=& \frac{\lambda ^{\frac{1}{2 \beta }}+1}{ \left( \lambda ^{\frac{2-\beta}{2 \beta }}-1\right)\left(\lambda ^{\frac{1}{2 \beta }}-1\right)},\\
	C &=& \frac{-2}{\left( \lambda ^{\frac{1-\beta}{2 \beta }}-1\right) \left(\lambda ^{\frac{1}{2 \beta }}-1\right)}\text{ } \cdot
\end{eqnarray}

\section{Stress derivation in the slender cone model}
\label{sec:StressCalculationInSlenderCone}

The stress state for the slender cone model is obtained by direct integration of the fluid force defined in Eq.~\eqref{eq:flin}, using Eq.~\eqref{eq:dz} for the diameter. The shear force and resulting bending moment read
\begin{equation}
	\tau(z) = \int_{z_0}^{z}f(z')\mathrm{d}z', \quad	M(z) = \int_{z_0}^{z}\tau(z')\mathrm{d}z',
\end{equation}
with $z_0 \geq 0$. The local non-dimensional skin stress reads
\begin{equation}
	\sigma(z) = \frac{32 M(z)}{\pi \Sigma_c d(z)^3}.
\end{equation}
The integration of these equations give Eq.~\eqref{eq:sig_inf} and Eq.~\eqref{eq:sig_trunc} depending on $z_0$.





\end{document}